\documentclass[aps,prd,reprint,groupedaddress,amsmath,amssymb]{revtex4-2}

\usepackage{graphicx} 
\usepackage{dcolumn}  
\usepackage{bm}       
\usepackage{hyperref} 
\usepackage{mathrsfs} 
\usepackage{xcolor, color, framed}
\newcommand{\trento}{T\raisebox{-.5ex}{R}ENTo}
\begin{document}

\title{System-size dependence of the $D^0$--$D_s^+$ flow splitting from early $D_s^+$ formation at $\sqrt{s_{NN}} = 5.36$~TeV}

\author{Hui Du}
\affiliation{School of Mathematics and Physics, China University of Geosciences, Wuhan 430074, China}

\author{Xiao-Wei Hao}
\affiliation{School of Mathematics and Physics, China University of Geosciences, Wuhan 430074, China}

\author{Wei Dai}
\email{corresponding author: weidai@cug.edu.cn} 
\affiliation{School of Mathematics and Physics, China University of Geosciences, Wuhan 430074, China}

\author{Jiaxing Zhao}
\affiliation{Helmholtz Research Academy Hesse for FAIR (HFHF), GSI Helmholtz Center for Heavy Ion Physics, Campus Frankfurt, 60438 Frankfurt, Germany}
\affiliation{Institut f\"ur Theoretische Physik, Johann Wolfgang Goethe-Universität, Max-von-Laue-Straße 1, D-60438 Frankfurt am Main, Germany}

\author{Ben-Wei Zhang}
\affiliation{Key Laboratory of Quark \& Lepton Physics (MOE) and Institute of Particle Physics, Central China Normal University, Wuhan 430079, China}

\author{Enke Wang}
\affiliation{State Key Laboratory of Nuclear Physics and Technology, Institute of Quantum Matter, South China Normal University, Guangzhou 510006, China}

\affiliation{Guangdong Basic Research Center of Excellence for Structure and Fundamental Interactions of Matter, Guangdong Provincial Key Laboratory of Nuclear Science, Guangzhou 510006, China}

\date{\today}

\begin{abstract}
We investigate the elliptic-flow splitting between prompt $D^0$ and $D_s^+$ mesons within a heavy-quark transport framework with sequential hadronization, in which $D_s^+$ forms at $1.2\,T_c$ and $D^0$ at $T_c$. We present predictions for the $p_T$-differential $v_2$ and $D_s^+/D^0$ yield ratio in O--O $0$--$20\%$ collisions at $\sqrt{s_{NN}} = 5.36$~TeV, where preliminary ALICE data are available. The sequential scenario reproduces the observed $v_2(D^0) > v_2(D_s^+)$ ordering and predicts an enhanced $D_s^+/D^0$ ratio at low $p_T$, whereas a simultaneous baseline yields the opposite ordering. Decomposing the hadronic splitting into its partonic components, we show that the $v_2$ ordering is driven by the late-stage flow accumulated by $D^0$-parent charm quarks during the $1.2\,T_c \to T_c$ interval, with hadronic rescattering essential to preserve the signal in small systems. A systematic scan across nine collision configurations spanning O--O and Pb--Pb centralities reveals a universal linear scaling between the hadronic splitting and the partonic flow increment accumulated during this window. This establishes the $D^0$--$D_s^+$ flow splitting as a hadronization chronometer of the QGP at $\sqrt{s_{NN}} = 5.36$~TeV.
\end{abstract}

\maketitle

\section{Introduction}

The production of heavy-flavor hadrons in ultrarelativistic heavy-ion collisions offers a calibrated probe of the hadronization mechanism in the quark-gluon plasma (QGP)~\cite{Andronic:2021erx}. Among open-charm mesons, the comparison between $D^0$ and $D_s^+$ is informative. The $D_s^+/D^0$ yield ratio is significantly enhanced in nucleus-nucleus collisions relative to proton-proton and proton-nucleus collisions, as established by the ALICE Collaboration in Pb--Pb measurements at $\sqrt{s_{NN}} = 2.76$ and $5.02$~TeV~\cite{ALICE:2016Ds,ALICE:2018Ds}. This enhancement arises because the abundant strangeness produced in the deconfined medium enables charm quarks to form $D_s^+$ via coalescence with thermal strange quarks, a channel that is presumed to be negligible in proton-proton collisions~\cite{ALICE:2016Ds,ALICE:2018Ds,Minissale:2023dct}. Coalescence-plus-fragmentation model calculations predict a monotonic increase in the $D_s^+/D^0$ ratio with increasing system size, from small to large collision systems~\cite{Minissale:2023dct,Liu:2024tfi}. A systematic scan across collision systems---from small O--O to large Pb--Pb---is essential to isolate the hadronization dynamics from system-specific initial geometry and lifetime effects.

Alongside the yield enhancement phenomenon, a characteristic ordering has emerged recently in the elliptic flow observable of charmed mesons: ALICE measurements in Pb--Pb collisions reveal $v_2(D_s^+) < v_2(D^0)$ at intermediate transverse momentum ($p_T$)~\cite{ALICE:2026zcz}. In most phenomenological approaches, all charmed hadrons are assumed to form on a single universal freeze-out hypersurface at the critical temperature $T_c$~\cite{Katz:2019qwv,Li:2021,Zhao:2024oma,Fries:2003kq,Greco:2003mm,He:2019vgs,He:2022ywp,Greco:2008coalescence,Cao:2015hia,Cao:2019iqs,Gossiaux:2009mk,He:2012df,Beraudo:2014boa,Zhao:2024ecc,Li:2021xbd,Song:2015sfa,Plumari:2017ntm,Das:2016llg}. Within this simultaneous-hadronization framework, however, coalescence kinematics favor the opposite ordering. Owing to the distinct thermal momentum distributions of strange and light quarks, the phase-space selection inherent to the recombination process preferentially picks charm quarks with larger $v_2$ to form $D_s^+$ mesons, leading to $v_{2}(D_s^+) > v_{2}(D^0)$ — in direct contradiction with the experimental observations. Resolving this ordering reversal is therefore a challenge for heavy-flavor hadronization theory.

An alternative paradigm is provided by the sequential hadronization scenario~\cite{Shi:2019tji,Zhao:2020rrk}. Based on the distinct binding energies of heavy-flavor hadrons obtained by solving the Dirac equation with an in-medium lattice potential~\cite{Shi:2013rga,Shi:2019tji}, we predicted that $D_s^+$ mesons are to form on an earlier hypersurface, presumably at $T \simeq 1.2\,T_c$, while $D^0$ and other charmed hadrons freeze out later at $T_c$. This time separation defines a finite late-stage evolution window, $\Delta\tau_{\rm had}$, during which the parent charm quarks destined for $D^0$ continue to accumulate momentum anisotropy from the expanding bulk medium. The natural consequence is $v_2(D_s^+) < v_2(D^0)$, with the splitting magnitude governed by the duration of the $1.2\,T_c \to T_c$ cooling phase. Our earlier analysis of Pb--Pb collisions at $30$--$50\%$ centrality identified this very signal and demonstrated quantitative agreement with ALICE data within the sequential framework~\cite{Xu:2025ivv}. While this earlier study established the viability of the sequential scenario in Pb--Pb, the microscopic mechanism---how the partonic flow increment during $\Delta\tau_{\rm had}$ translates into the hadronic $v_2$ reversal, and how this picture survives when the time window is compressed---remains to be fully dissected.

O--O collisions, with their far shorter QGP lifetime, provide a particularly stringent test of whether the sequential-hadronization picture holds across system sizes: they probe whether signatures of distinct hadronization timescales survive when the late-stage evolution window is severely compressed~\cite{ALICE:2025strangeness,Giacalone:2024luz}. The recent availability of ALICE Run-3 O--O collision data at $\sqrt{s_{NN}} = 5.36$~TeV~\cite{alice_oo_charm_v2_sqm2026} now makes this test possible. The preliminary data reveal that $v_2(D_s^+) < v_2(D^0)$ indeed persists in O--O, albeit with a markedly smaller magnitude than in Pb--Pb. These observations motivate a comprehensive investigation that dissects the microscopic origin of the $v_2$ ordering and maps its dependence on system size.

The present work is driven by two interconnected goals. \textit{First}, we aim to uncover the microscopic mechanism by which sequential hadronization generates the observed $v_2(D^0) > v_2(D_s^+)$ ordering, and to contrast it with the reversed baseline expected from simultaneous coalescence kinematics. \textit{Second}, by systematically scanning collision systems from O--O to Pb--Pb, we identify the experimental conditions---system size and centrality---under which the sequential-hadronization signal is maximized. Toward these ends, we present in Sec.~\ref{sec:results} predictions for the $p_T$-differential $v_2$ and $D_s^+/D^0$ yield ratio in O--O $0$--$20\%$ collisions, contrast the sequential and simultaneous scenarios against preliminary ALICE data, and compare the splitting across O--O and Pb--Pb systems. In Sec.~\ref{sec:decomposition}, we decompose the hadronic flow splitting into its partonic components by examining the parent charm-quark $v_2$ at the $1.2\,T_c$ and $T_c$ hypersurfaces in both O--O and Pb--Pb systems, isolating the roles of coalescence kinematics, fragmentation mixing, and hadronic rescattering. Finally, in Sec.~\ref{sec:scaling}, we perform a systematic scan across nine collision configurations spanning O--O and Pb--Pb and test whether the hadronic splitting obeys a universal linear scaling with the partonic flow increment accumulated during the sequential-hadronization window, establishing the $D^0$--$D_s^+$ splitting as a chronometer of the late-stage QGP evolution.
\section{Theoretical Framework}
\label{sec:method}

In this section, we present the multi-stage theoretical framework employed to simulate the heavy-quark evolution and hadronization. The framework integrates initial state generation, (3+1)D viscous hydrodynamic expansion, an improved Langevin transport approach, and a hybrid hadronization scheme.

\subsection{Initial conditions of the O--O system and hydrodynamic description}

The space-time evolution of the bulk medium is simulated by coupling an initial-state generator with (3+1)D relativistic viscous hydrodynamics. For the initial geometry, we employ the \trento\ model~\cite{Moreland:2014oya}. For the O--O system, the nucleonic density profiles are taken from the Projected Generator Coordinate Method (PGCM) configurations in Ref.~\cite{Giacalone:2024luz} to account for the intrinsic $\alpha$-clustering structure of the oxygen nucleus.

Within the \trento\ framework, the initial entropy density is deposited proportionally to the reduced thickness, constructed from the participant nuclear thickness functions. Each thickness function is modeled as a superposition of fluctuating Gaussian nucleons with an effective width $w$, and sub-nucleonic multiplicity fluctuations are incorporated by independently sampling the weight of each nucleon from a Gamma distribution with shape parameter $k$. We set $k = 1.5$ and $w = 0.43$~fm. 

\begin{figure}[htbp]
    \centering
    \includegraphics[width=1.05\linewidth]{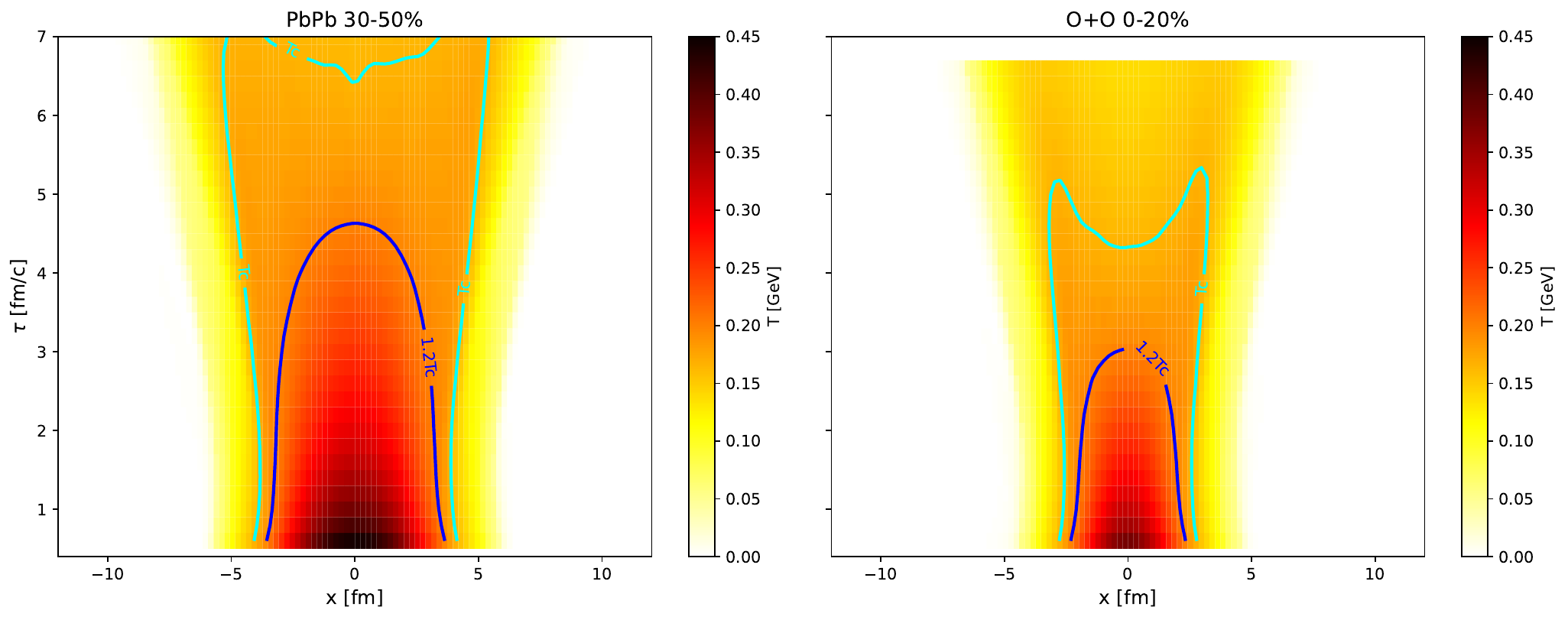}
    \caption{Spatial profiles of the isothermal hypersurfaces at $T = 1.2\,T_c$ (blue) and $T = T_c$ (cyan) in the $\tau$--$x$ plane, evaluated at mid-rapidity for O--O $0$--$20\%$ (right) and Pb--Pb $30$--$50\%$ (left) collisions. The separation between the two hypersurfaces defines the late-stage evolution window probed by the sequential hadronization scenario.}
    \label{fig:liu}
\end{figure}

The hydrodynamic expansion of the QGP is described by the (3+1)D CLVisc framework~\cite{Pang:2018zzo,Wu:2021fjf}, starting at an initial proper time $\tau_0 = 0.6$~fm/$c$. The equation of state is taken from the HotQCD2014 lattice result~\cite{HotQCD:2014kol}. The medium's dissipative properties are parametrized by temperature-dependent shear and bulk viscosities. For the shear viscosity, $\eta/s(T)$ maintains a minimum of $0.08$ in the high-temperature phase and increases linearly below $T_c = 0.154$~GeV. For the bulk viscosity, $\zeta/s(T)$ is parametrized as a skewed double-exponential peak centered at $T_c$, whose shape and magnitude were determined through Bayesian model-to-data comparisons~\cite{Bernhard:2016tnd,Ryu:2015vwa}. The hydrodynamic evolution continues until the system cools below the freeze-out hypersurface, which defines the spacetime boundary at which heavy-flavor hadronization takes place. In the sequential hadronization scenario, however, $D_s^+$ mesons are formed on an earlier hypersurface at $T \simeq 1.2\,T_c$, while $D^0$ mesons freeze out at $T_c$. Figure~\ref{fig:liu} illustrates this geometry by comparing the isothermal contours at $T = 1.2\,T_c$ and $T = T_c$ in the $\tau$--$x$ plane for both O--O and Pb--Pb collisions. The figure highlights two distinct sources of system-size dependence: the overall space extent differs between the two collision systems, and within each system the time interval between the two hypersurfaces provides the late-stage evolution window that drives the $D_s^+$--$D^0$ flow splitting, as will be quantified in Secs.~\ref{sec:results}--\ref{sec:scaling}.

\subsection{Heavy quark transport}
The space-time evolution of charm quarks within the dynamically expanding QGP is described by a Langevin approach that accounts for both quasi-elastic scattering and radiative energy loss~\cite{Cao:2013ita,Cao:2015hia,Dai:2018mhw}. To initialize this evolution, the space coordinates of charm quarks are sampled according to the binary nucleon-nucleon collision ($N_{\rm coll}$) profile, while their initial momentum spectra are taken from Fixed-Order-plus-Next-to-Leading-Logarithms (FONLL) calculations~\cite{Cacciari:2012ny}. To account for cold nuclear matter effects in the O--O system, the EPPS21 nuclear parton distribution functions (nPDFs)~\cite{Eskola:2021nhw} are explicitly incorporated, which yield a sizable shadowing suppression of the initial charm-quark production cross section at low $p_T$. The numerical simulation tracks these heavy quarks as they propagate through the hydrodynamic cells provided by the $\mathtt{CLVisc}$ background. For a charm quark with energy $E = \sqrt{p^2 + M^2}$, the updates of its space coordinates $x_j$ and momentum $p_j$ in each time step $dt$ are described by the following set of equations~\cite{Wang:2019xey,Wang:2020qwe,Wang:2020ukj,Dai:2018mhw,Xue:2025hrw}:
\begin{align}
dx_j &= \frac{p_j}{E} dt, \\
dp_j &= -\Gamma(p) p_j dt + \sqrt{\kappa (p)dt}\rho_j - \vec{p}_{g},
\end{align}
where the terms on the right-hand side of the momentum update represent the deterministic drag force, the stochastic thermal force, and the momentum change due to radiated gluons, respectively.

In this formulation, $\Gamma(p)$ is the drag coefficient and $\kappa$ denotes the momentum diffusion coefficient. The stochastic variable $\rho_j$ represents an independent Gaussian noise following a normal distribution with zero mean and unit variance, satisfying $\langle \rho_j \rangle = 0$ and $\langle \rho_j(t) \rho_k(t') \rangle = \delta_{jk} \delta_{tt'}$. The drag and diffusion coefficients are intrinsically linked through the fluctuation-dissipation relation, $\kappa = 2\Gamma ET$~\cite{Kubo:1966fyg}, which is further related to the spatial diffusion coefficient of charm quarks via $D_s = 2T^2/\kappa$. Guided by 2+1 flavor lattice QCD calculations~\cite{Altenkort:2023oms,Petreczky:2010yn} and taking the parameters from previous Pb--Pb studies~\cite{Xu:2025ivv,Xue:2025hrw} as a reference benchmark, we vary the spatial diffusion coefficient around a comparable baseline, setting $2\pi T D_s = 2.1 \pm 0.4$ for the O--O system, and treat it as temperature-independent.

The term $-\vec{p}_g$ accounts for the recoil momentum exerted on the heavy quark during each gluon radiation event. The radiation process is implemented based on the medium-induced gluon distribution within the higher-twist approach~\cite{Guo:2000nz,Zhang:2003wk,Zhang:2004qm,Majumder:2009ge,Cao:2013ita}:
\begin{equation}
\frac{dN_g}{dx dk_\perp^2 dt} = \frac{2\alpha_s C_s P(x)\hat{q} }{\pi k_\perp^4} \sin^2\left( \frac{t - t_{i}}{2\tau_{f}} \right) \left[ \frac{k_\perp^2}{k_\perp^2 + (xM)^2} \right]^4,
\end{equation}
where $x$ and $k_\perp$ denote the energy fraction and transverse momentum of the emitted gluon, $\alpha_s$ is the strong coupling constant, $C_s$ is the quadratic Casimir in color representation, $P(x)$ is the vacuum splitting function~\cite{Deng:2009ncl}, and $\tau_f = 2Ex(1-x)/[k_\perp^2 + (xM)^2]$ is the formation time. The jet quenching parameter is parametrized as $\hat{q} = \hat{q}_0 (T^3/T_c^3) (p_{\mu}u^{\mu}/p^0)$. Guided by previous Pb--Pb studies~\cite{Xu:2025ivv,Xie:2019oxg}, we set the quenching strength for the O--O system to $\hat{q}_0 = 1.4 \pm 0.3 \text{ GeV}^2/\text{fm}$, where the variation range explicitly accounts for systematic uncertainties when extrapolating the transport framework to smaller collision volumes. For Pb--Pb collisions, the central values of the transport parameters are taken from our previous study~\cite{Xu:2025ivv,Xue:2025hrw}, whereas the uncertainty bands are shown only for the O--O system.

After hadronization, the formed $D$ mesons undergo rescattering in the hadronic phase. This stage is also described by a Langevin equation, with a temperature-dependent spatial diffusion coefficient for $D$-meson interactions with the light-flavor hadron gas~\cite{He:2012df,He:2019tik,Torres-Rincon:2021yga}. The hadronic rescattering continues until the system reaches a fixed kinetic freeze-out temperature of $137$~MeV~\cite{Pang:2018zzo}.

\subsection{Hadronization}
This section describes the hadronization of charm quarks into charmed hadrons via a hybrid coalescence-plus-fragmentation model~\cite{Fries:2003kq,Greco:2003mm,Fries:2008hs}. Charm quarks first hadronize through coalescence with thermal light and strange quarks at the hadronization hypersurface; those that do not coalesce are subsequently converted via fragmentation. The momentum distribution of hadrons produced from quark coalescence is given by
\begin{eqnarray}
\label{eq:coal}
{dN\over d^3{\bf P}}&=&g_H \int \prod_{i=1}^n{d^3x_id^3p_i\over (2\pi)^3E_i} f_i({\bf x}_i,{\bf p}_i)\\
&\times& W_h({\bf x}_1,\cdots,{\bf x}_n, {\bf p}_1,\cdots, {\bf p}_n)\, \delta^{(3)}\left({\bf P}-\sum_{i=1}^k{\bf p}_i\right),\nonumber
\end{eqnarray}   
where $i$ indexes the constituent quarks and $g_H$ is the spin-color degeneracy factor. The phase-space distribution $f_i(x_i, p_i)$ of each constituent quark is evaluated at the hadronization hypersurface. For thermal light quarks, a thermal distribution is taken, $f_q(x_i, p_i) = N_q / (e^{u_{\mu}p_i^{\mu}/T} + 1)$ with degeneracy $N_q = 6$.

The Wigner function $W_h$ represents the probability density for the parton pair to form a specific hadron state. Taking charmed mesons as an example, the phase-space Wigner functions for the specific wave states involved in this model read~\cite{Zhao:2018jlw,Zhao:2023nrz,Zhao:2025cnp}:
\begin{align}
W_{1S}(\mathbf{r}, \mathbf{p}) =& 8e^{-\xi}, \notag \\
W_{1P}(\mathbf{r}, \mathbf{p}) = &\frac{8}{3}e^{-\xi}(2\xi - 3), \notag \\
W_{1D}(\mathbf{r}, \mathbf{p}) = &\frac{8}{15}e^{-\xi}(15 + 4\xi^2 - 20\xi + 8\eta), \notag \\
W_{2S}(\mathbf{r}, \mathbf{p}) =& \frac{8}{3}e^{-\xi}(3 + 2\xi^2 - 4\xi - 8\eta), \notag \\
W_{2P}(\mathbf{r}, \mathbf{p}) = &\frac{8}{15}e^{-\xi}(-15 + 4\xi^3 - 22\xi^2 + 30\xi \notag \\
&- 8(2\xi - 7)\eta),
\end{align}
where we define $\xi \equiv r^2/\sigma^2 + p^2\sigma^2$ and $\eta \equiv p^2r^2 - (\mathbf{p} \cdot \mathbf{r})^2$. Here, $\mathbf{r}$ and $\mathbf{p}$ denote the relative coordinate and relative momentum of the two constituent quarks in their center-of-mass (CM) frame. For charmed mesons, the relative momentum is given by:
\begin{equation}
\mathbf{p} = \frac{m_1 \mathbf{p}_c - m_c \mathbf{p}_1}{m_1 + m_c}.
\end{equation}
The Gaussian width parameter $\sigma$ encoding the spatial extension of the wave function is determined by the average radius of the meson:
\begin{equation}
\langle r \rangle = \int \frac{d^3r d^3p}{(2\pi)^3} W_h(\mathbf{r}, \mathbf{p}) r.
\end{equation}
The averaged radius of the ground state for charmed mesons and baryons is calculated by solving the Dirac equation with a lattice quark potential \cite{Shi:2013rga,Shi:2019tji}. To simplify the subsequent calculations, we integrate out the coordinate-space degrees of freedom from the Wigner function and retain only the momentum-space distribution in this study. The coalescence probabilities of various charmed hadrons are evaluated in a thermal box at a given temperature using a formula analogous to Eq.~\eqref{eq:coal}, with the charm-quark distribution approximated by a delta function $\delta^{(3)}({\bf p}-{\bf p}_c)$.

Based on the distinct binding energies of heavy-flavor hadrons, we implement a sequential hadronization scenario governed by the temperature ordering\cite{Shi:2019tji}:
\begin{equation}
T_{D_s} \simeq 1.2T_c > T_{D^0}, T_{\Omega_c}, T_{\Xi_c}, T_{\Lambda_c} \simeq T_c.
\end{equation}
In this approach, when a charm quark with momentum $p$ (determined by the Langevin evolution) reaches the earlier coalescence hypersurface at $T_{D_s}$, its probability $P_{D_s}$ to form a $D_s^+$ meson is computed from the coalescence integral, Eq.~(\ref{eq:coal}), evaluated at that hypersurface. If coalescence occurs, the charm quark hadronizes into $D_s^+$ and exits the partonic evolution. If not, it continues to transport through the QGP until reaching the $T_c$ hypersurface, where it undergoes a second coalescence stage into other charmed hadrons, with the available probability rescaled by $1 - P_{D_s}$.

After these two coalescence stages at $T_{D_s}$ and $T_c$, if the charm quark still survives, it hadronizes via fragmentation. The momentum distribution of the final-state hadron is then governed by the Peterson fragmentation function\cite{Peterson:1982ak}:
\begin{equation}
\mathscr{D}_{c \to H}(z) \propto \frac{1}{z \left( 1 - \frac{1}{z} - \frac{\epsilon}{1-z} \right)^2},
\end{equation}
where $z$ is the momentum fraction of the charmed hadron fragmented from the parent charm quark. We adopt $\epsilon = 0.01$ for charmed mesons and $\epsilon = 0.02$ for charmed baryons\cite{Das:2016llg}. The relative fragmentation fractions of a charm quark into various specific charmed hadrons are constrained by experimental data from $p+p$ collisions~\cite{Lisovyi:2015uqa}.
\section{Predictions for O--O collisions and system-size comparison}
\label{sec:results}

We present theoretical predictions for prompt $D^0$ and $D_s^+$ production in 0--20\% central O--O collisions at $\sqrt{s_{NN}} = 5.36$~TeV~\cite{Minissale:2020bif,Plumari:2017ntm}, and compare the sequential- and simultaneous-hadronization scenarios against preliminary ALICE data.

\begin{figure}[t]
    \centering
    \includegraphics[width=1.0\columnwidth]{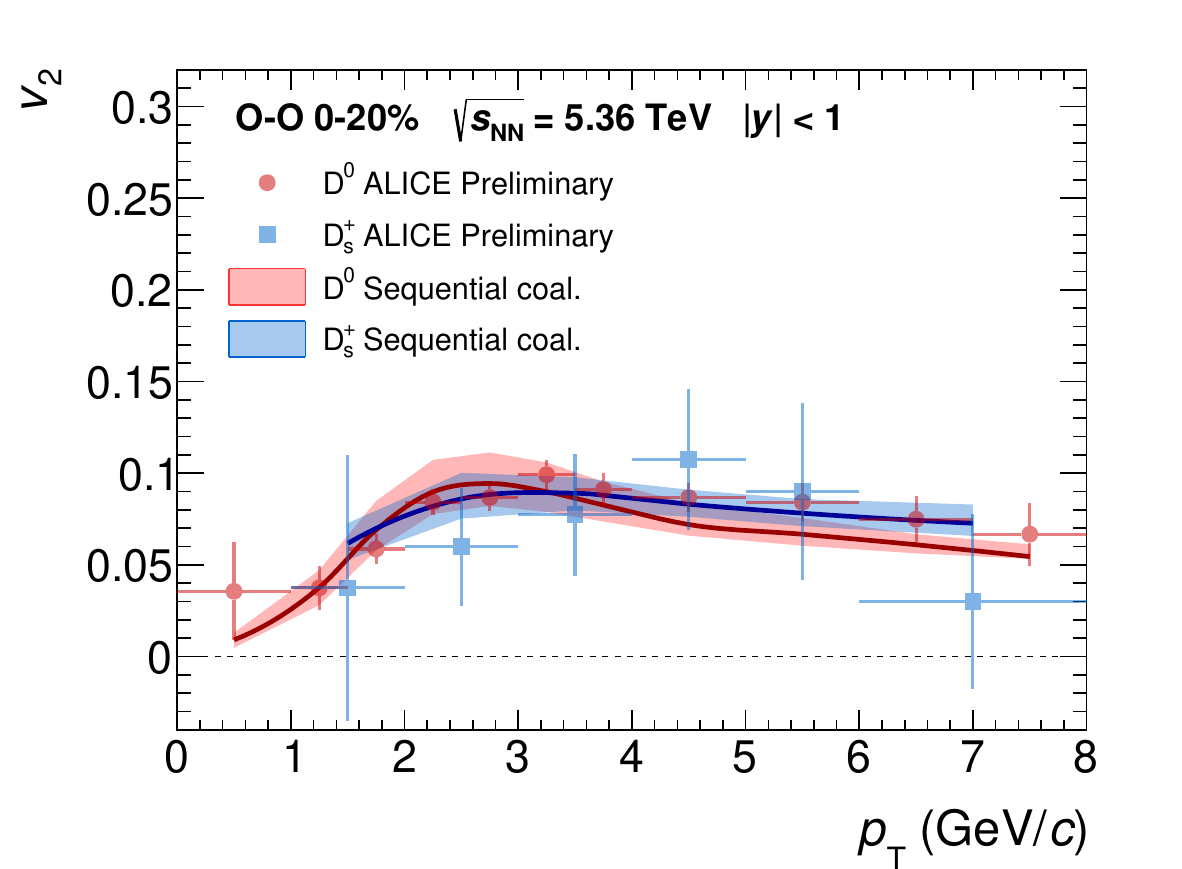}
    \vspace{0.0cm}
    \includegraphics[width=1.0\columnwidth]{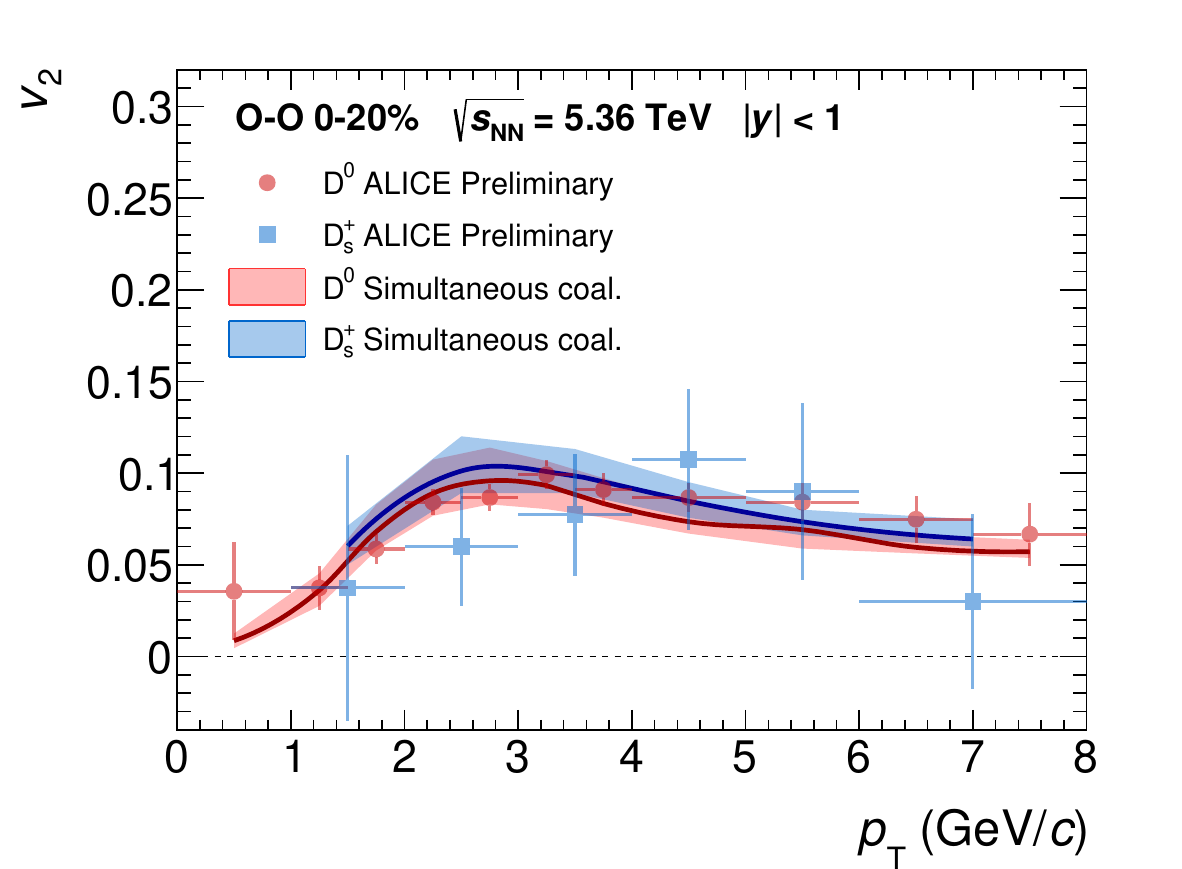}
    \caption{Elliptic flow $v_2$ of prompt $D^0$ and $D_s^+$ as a function of $p_T$ in 0--20\% central O--O collisions at $\sqrt{s_{NN}} = 5.36$~TeV. Bands reflect variations of $2\pi T D_s = 2.1 \pm 0.4$ and $\hat{q}_0 = 1.4 \pm 0.3$~GeV$^2$/fm; solid lines show central values. Upper and lower panels correspond to the sequential and simultaneous hadronization scenarios, respectively.}
    \label{fig:theory_vs_data_comparison}
\end{figure}

Figure~\ref{fig:theory_vs_data_comparison} (upper panel) presents the $p_T$-differential $v_2$ of prompt $D^0$ and $D_s^+$ mesons calculated within the sequential hadronization scenario. The calculations reproduce the $v_2(D^0) > v_2(D_s^+)$ ordering observed in the data at intermediate $p_T$, with the theoretical uncertainty bands---reflecting variations of $2\pi T D_s = 2.1 \pm 0.4$ and $\hat{q}_0 = 1.4 \pm 0.3$~GeV$^2$/fm---in good overall agreement with the measurements. In the sequential scenario, $D_s^+$ mesons freeze out on the earlier $T \simeq 1.2\,T_c$ hypersurface with less accumulated momentum anisotropy, while $D^0$-parent charm quarks continue to acquire $v_2$ throughout the $1.2\,T_c \to T_c$ cooling phase, naturally producing the correct ordering even within the short-lived QGP of O--O collisions.

\begin{figure}[t]
    \centering
    \includegraphics[width=\linewidth]{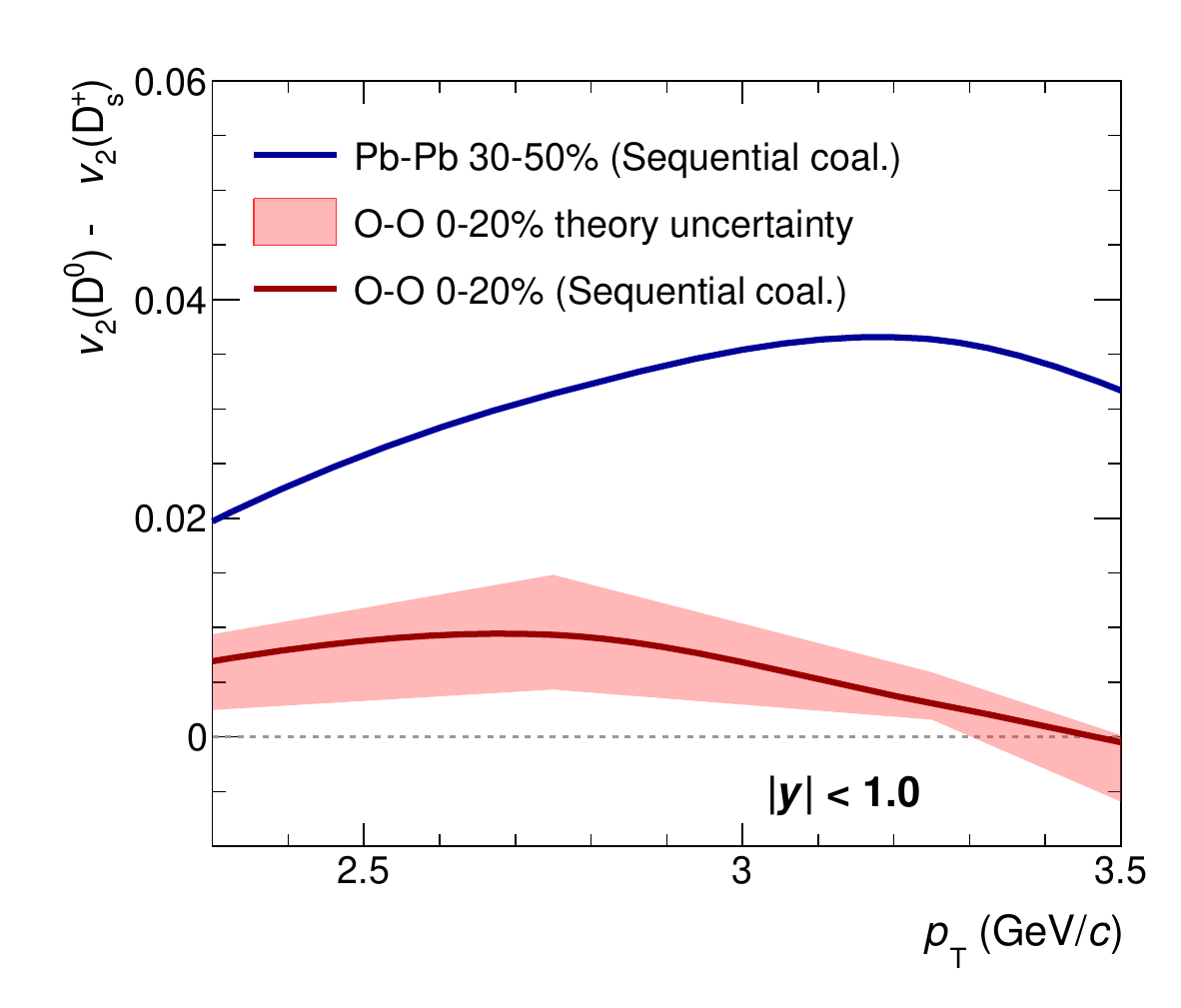}
    \caption{Hadronic flow splitting $\Delta v_2 = v_2(D^0) - v_2(D_s^+)$ as a function of $p_T$ within the sequential hadronization framework, comparing 0--20\% O--O at $\sqrt{s_{NN}} = 5.36$~TeV and 30--50\% Pb--Pb at $\sqrt{s_{NN}} = 5.02$~TeV collisions. The solid line and shaded band show the central prediction and theoretical uncertainty for O--O.}
    \label{fig:2}
\end{figure}

Figure~\ref{fig:theory_vs_data_comparison} (lower panel) presents the simultaneous scenario, in which both species are forced to form at $T_c$. While the $D^0$ $v_2$ remains well described, the predicted $D_s^+$ $v_2$ deviates noticeably from the data, yielding an inverted $v_2(D_s^+) > v_2(D^0)$ ordering at $p_T \lesssim 4$~GeV/$c$. Without the earlier $D_s^+$ freeze-out, the coalescence phase space preferentially selects charm quarks with larger $v_2$ to form $D_s^+$ mesons---a kinematic bias that drives the reversed ordering. The fact that only the sequential scenario reproduces the data indicates that the time separation between the two hadronization hypersurfaces is the essential ingredient for describing heavy-flavor flow in O--O collisions.
If the $v_2$ splitting is controlled by the duration of the $1.2\,T_c \to T_c$ window, its magnitude should be sensitive to the fireball lifetime, which varies with system size. We quantify this dependence in Fig.~\ref{fig:2}, which compares the hadronic flow splitting $\Delta v_2(p_T) \equiv v_2(D^0) - v_2(D_s^+)$ from the sequential scenario for O--O (0--20\%) and Pb--Pb (30--50\%) collisions. The peak $\Delta v_2$ in Pb--Pb is substantially larger than in O--O, reflecting the extended QGP lifetime in the heavier system. In the shorter-lived O--O system, the late-stage accumulation is compressed, yielding a reduced but still positive $\Delta v_2$.

In this comparison, the Pb--Pb result uses the central values of the transport parameters from our previous study~\cite{Xu:2025ivv,Xue:2025hrw}, while an uncertainty band is introduced for O--O to account for extrapolation of the transport coefficients to a smaller collision system. The positive $\Delta v_2$ persists across the explored parameter range, confirming that the sequential signal survives even under the compressed conditions of small systems. This sensitivity to the macroscopic space-time scale demonstrates that $\Delta v_2$ directly reflects the late-stage QGP evolution.

\begin{figure}[htbp]
    \centering
    \includegraphics[width=1.0\linewidth]{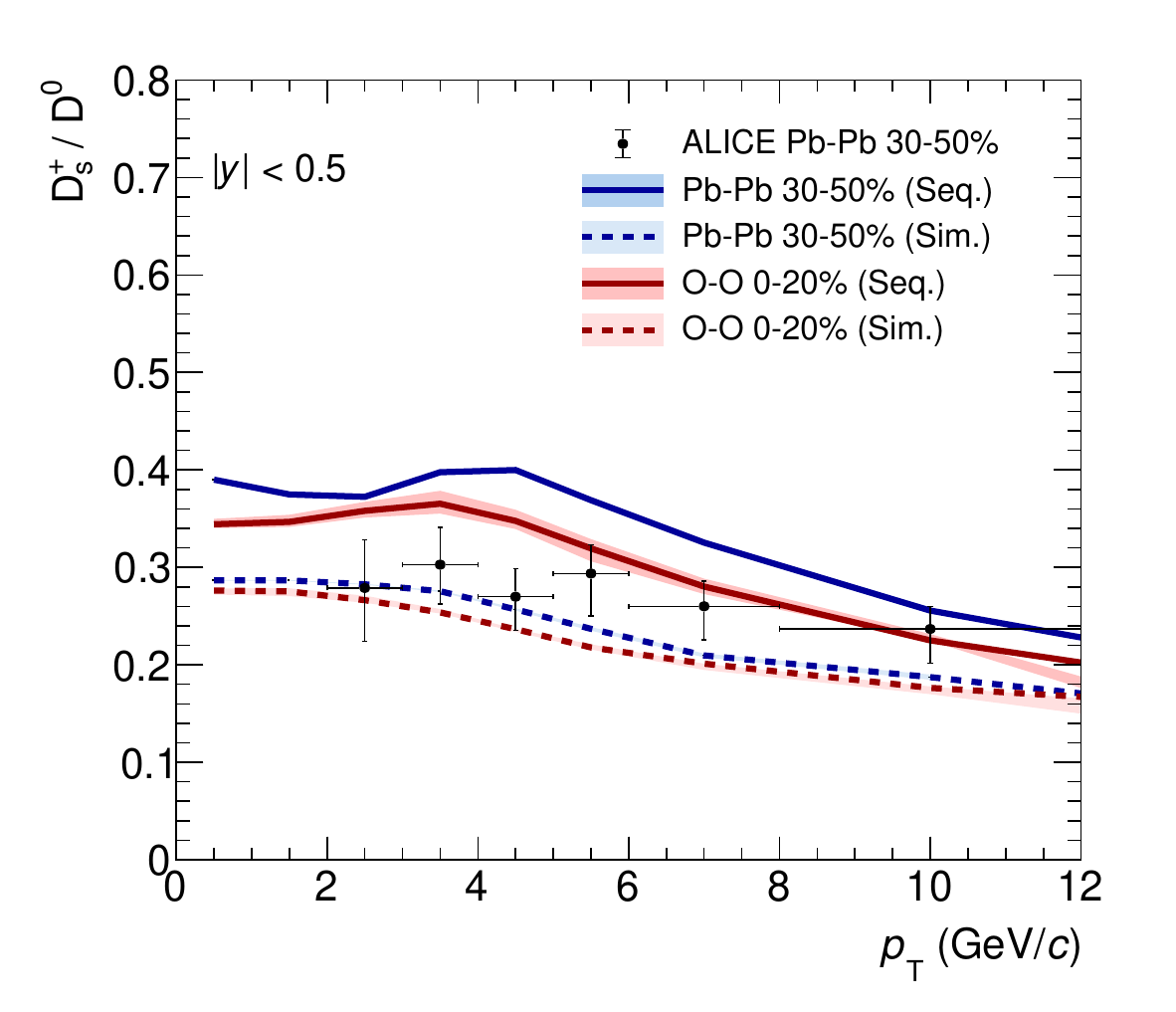}
    \caption{$D_s^+/D^0$ yield ratio as a function of $p_T$ in 0--20\% O--O at $\sqrt{s_{NN}} = 5.36$~TeV and 30--50\% Pb--Pb at $\sqrt{s_{NN}} = 5.02$~TeV, shown for both the sequential and simultaneous hadronization scenarios.}
    \label{fig:ds/d0}
\end{figure}

Figure~\ref{fig:ds/d0} shows the $D_s^+/D^0$ yield ratio as a function of $p_T$ for O--O and Pb--Pb collisions under both hadronization scenarios. In both systems, the sequential scenario yields a larger $D_s^+/D^0$ ratio than the simultaneous scenario at low $p_T$. Comparing the two systems, the $D_s^+/D^0$ ratio is systematically higher in Pb--Pb than in O--O across the full $p_T$ range.

The enhancement of the $D_s^+/D^0$ ratio in the sequential scenario reflects the higher coalescence probability for $D_s^+$ on the earlier $1.2\,T_c$ hypersurface, where the strange-quark phase-space density is larger than at $T_c$. The systematic increase from O--O to Pb--Pb is driven by the stronger partonic energy loss in the heavier system, which shifts the parent charm-quark spectrum to lower $p_T$ and thereby enhances the coalescence probability at the $1.2\,T_c$ hypersurface. Together with the $v_2$ results discussed above, these trends demonstrate that the hadronization scheme and the collision system size jointly govern both the kinematic and chemical properties of open-charm hadrons.

\section{Partonic Origin of the Flow Splitting}
\label{sec:decomposition}

To identify the partonic origin of the $D^0$--$D_s^+$ flow splitting, we examine the dynamics at the level of the parent charm quarks prior to hadronization. The key quantity is the time separation between the two formation hypersurfaces, $\Delta\tau_{\rm had} = \langle \tau_{D^0} \rangle - \langle \tau_{D_s^+} \rangle$, extracted from the sequential hadronization scenario.
\begin{figure}[t]
    \centering
    \includegraphics[width=1.05\linewidth]{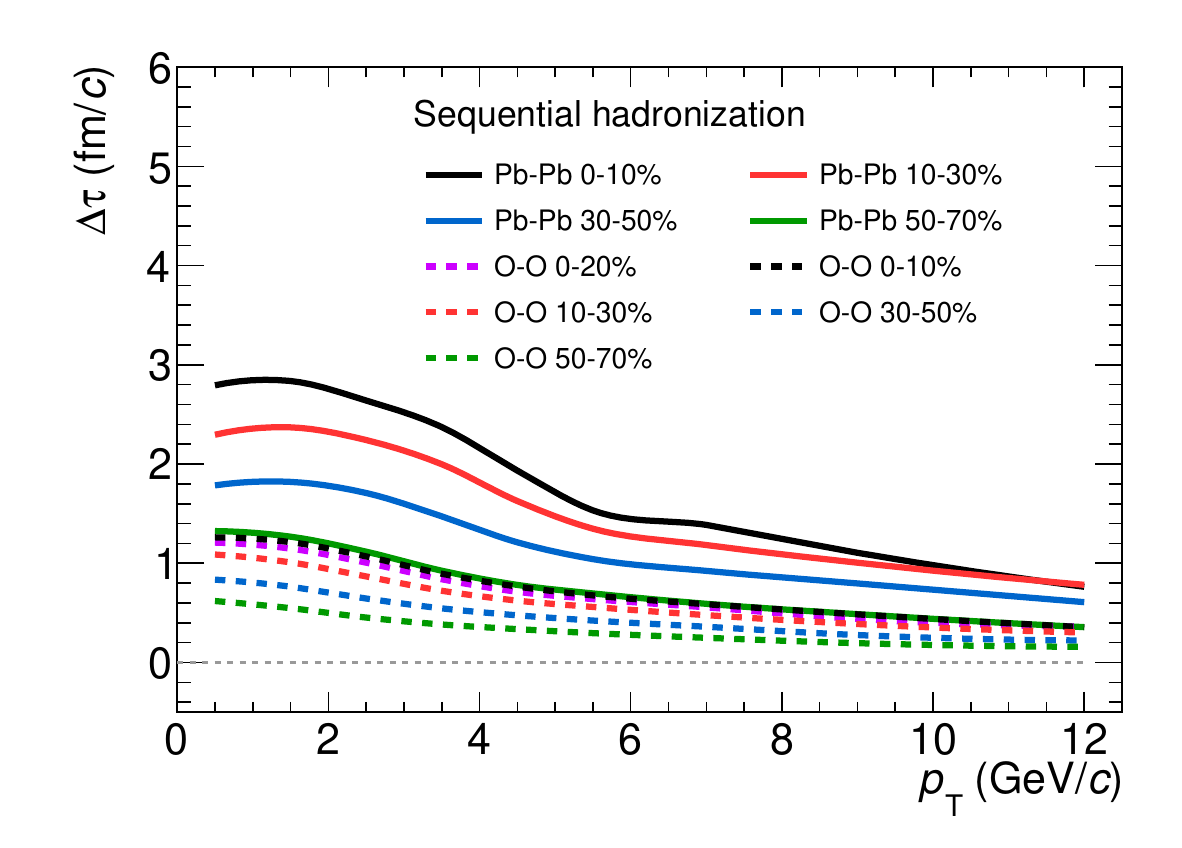}
    \caption{Average hadronization time difference $\Delta\tau_{\rm had} = \langle \tau_{D^0} \rangle - \langle \tau_{D_s^+} \rangle$ as a function of $p_T$, extracted from the sequential hadronization scenario across nine collision configurations in Pb--Pb and O--O systems.}
    \label{fig:tau_diff}
\end{figure}

Figure~\ref{fig:tau_diff} illustrates the transverse-momentum ($p_T$) dependence of $\Delta\tau_{\rm had}$, evaluated across nine distinct configurations encompassing various centrality classes in both Pb--Pb and O--O collisions. The calculations reveal a pronounced sensitivity to the system size, event centrality, and $p_T$. Specifically, $\Delta\tau_{\rm had}$ is maximized at low $p_T$ and exhibits a monotonic decrease toward higher $p_T$. Macroscopically, this time difference reflects the overall lifetime of the dynamically expanding fireball; in midcentral Pb--Pb collisions, $\Delta\tau_{\rm had}$ reaches $2$--$3$~fm/$c$ at low $p_T$, whereas the rapid expansion characteristic of the smaller O--O systems significantly compresses this temporal window. Physically, this finite $\Delta\tau_{\rm had}$ corresponds to the macroscopic cooling interval between the $T = 1.2\,T_c$ and $T = T_c$ hypersurfaces. During this extended late-stage evolution, the parent charm quarks of the $D^0$ mesons continue to scatter within the hydrodynamically expanding medium, thereby accumulating additional momentum anisotropy. We demonstrate that this late-time partonic flow acquisition is the fundamental driver of the observed hadronic flow splitting, $\Delta v_2 \equiv v_2(D^0) - v_2(D_s^+)$.
\begin{figure}[t]
    \centering
    \includegraphics[width=1.0\columnwidth]{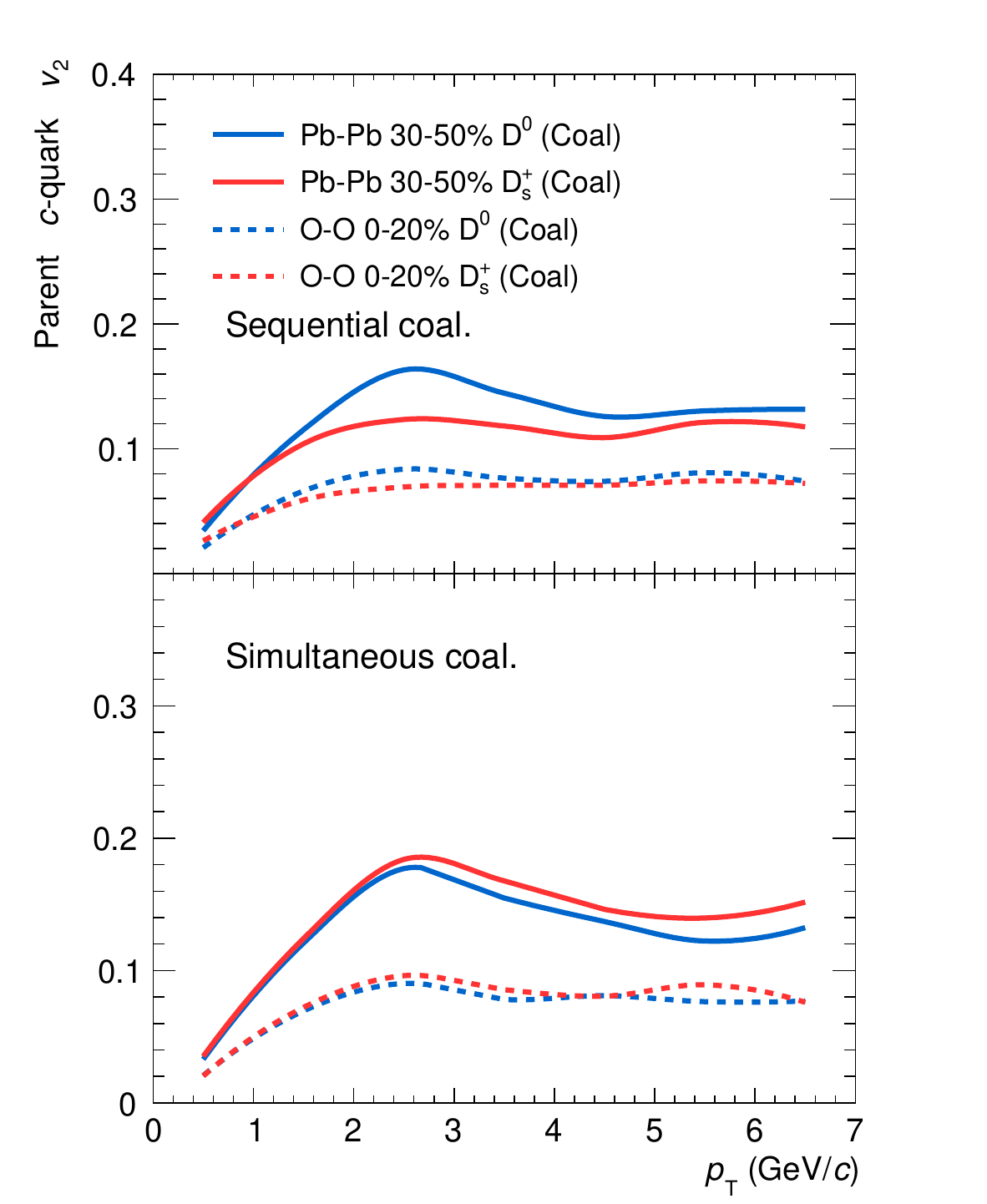}
    \caption{Elliptic flow $v_{2,c}$ of parent charm quarks  immediately before hadronization as a function of partonic $p_T^c$, for Pb--Pb 30--50\% (solid) and O--O 0--20\% (dotted) collisions. Upper and lower panels correspond to the sequential and simultaneous hadronization scenarios, respectively. Red and blue curves correspond to charm quarks destined for $D_s^+$ and $D^0$, respectively.}
    \label{fig:partonic_v2}
\end{figure}

Figure~\ref{fig:partonic_v2} displays the partonic $v_2$ of parent charm quarks ($v_{2,c}$) in the coalescence channel evaluated just before hadronization. In the sequential scenario (upper panel), $D_s^+$ parent charm quarks (red) freeze out earlier at $1.2\,T_c$ with less accumulated $v_2$, while those destined for $D^0$ (blue) continue to gain anisotropy down to $T_c$, yielding $v_{2,c}(D^0) > v_{2,c}(D_s^+)$ in both Pb--Pb and O--O. In the simultaneous limit (lower panel), coalescence phase-space selection instead produces $v_{2,c}(D_s^+) > v_{2,c}(D^0)$; the sequential time separation must therefore supply sufficient partonic flow to overcome this inverted baseline~\cite{Beraudo:2014boa,Beraudo:2017gxw}. 

Beyond the coalescence channel, a fraction of charm quarks hadronize via fragmentation. The partonic $v_2$ inherited through fragmentation is substantially smaller than that from coalescence, and is identical between the sequential and simultaneous scenarios because fragmentation occurs at the same $T_c$ hypersurface in both schemes. The parent charm $v_2$ is therefore a mixture of the two contributions. Since $D_s^+$ has a higher final state coalescence contribution fraction than $D^0$, the total parent-charm $v_2$ of $D^0$ receives a larger suppression from the low-$v_2$ fragmentation component than that of $D_s^+$. Due to this complication, the parton-level $v_2$ gap between $D^0$ and $D_s^+$ parent charm quarks is further narrowed in Pb--Pb collisions. In O--O collisions, where the $1.2\,T_c \to T_c$ window is severely compressed, the contribution mixing can even invert the partonic ordering, risking $v_{2,c}(D_s^+) > v_{2,c}(D^0)$ at the parton level. The hadronic rescattering phase, in which $D^0$ acquires a larger $v_2$ boost than $D_s^+$, is therefore essential to restore the $v_2(D^0) > v_2(D_s^+)$ ordering in small systems.

This rescattering phase at the hadron level means: $D^0$ mesons, with their larger hadronic cross sections, acquire an additional positive $v_2$ boost relative to $D_s^+$. Since this hadronic amplification is proportional to the partonic flow increment acquired during $1.2\,T_c \to T_c$, it acts as a dynamical amplifier that steepens the correlation between the final hadronic splitting and the late-stage partonic accumulation, as will be quantified in Sec.~\ref{sec:scaling}. Taking into account all these effects, we choose the $D^0$-parent increment rather than the difference $v_{2,c}(D^0) - v_{2,c}(D_s^+)$ because the latter is contaminated by the coalescence phase-space baseline and by species-dependent fragmentation mixing, whereas the former isolates the flow accumulated solely during the sequential-hadronization window. In the subsequent mechanism study, we therefore adopt the partonic $v_2$ increment $\Delta v_{2,c}$ accumulated by $D^0$ parent charm quarks during the $1.2\,T_c \to T_c$ interval as a proxy for the effective evolution time difference between the two hadronization hypersurfaces.

\section{Universal Scaling of Late-stage Evolution}
\label{sec:scaling}

The preceding analysis establishes that $\Delta\tau_{\rm had}$ is the primary driver behind the $D^0$--$D_s^+$ flow splitting. To test whether this connection is universal, we perform a systematic scan across nine LHC collision configurations at $5.36 $~TeV---spanning Pb--Pb and O--O centralities---all evaluated under the sequential hadronization scenario. We examine whether the hadronic flow splitting and the partonic flow increment obey a common scaling relation across this wide range of system sizes and initial geometries. All observables in this section are integrated over the transverse momentum window $2.5 < p_T^{\rm had} < 3.0$~GeV/$c$, which corresponds to the intermediate-$p_T$ region where coalescence is the dominant hadronization channel and the $D^0$--$D_s^+$ flow splitting is most pronounced.
\begin{figure}[t]
    \centering
    \includegraphics[width=1.08\columnwidth]{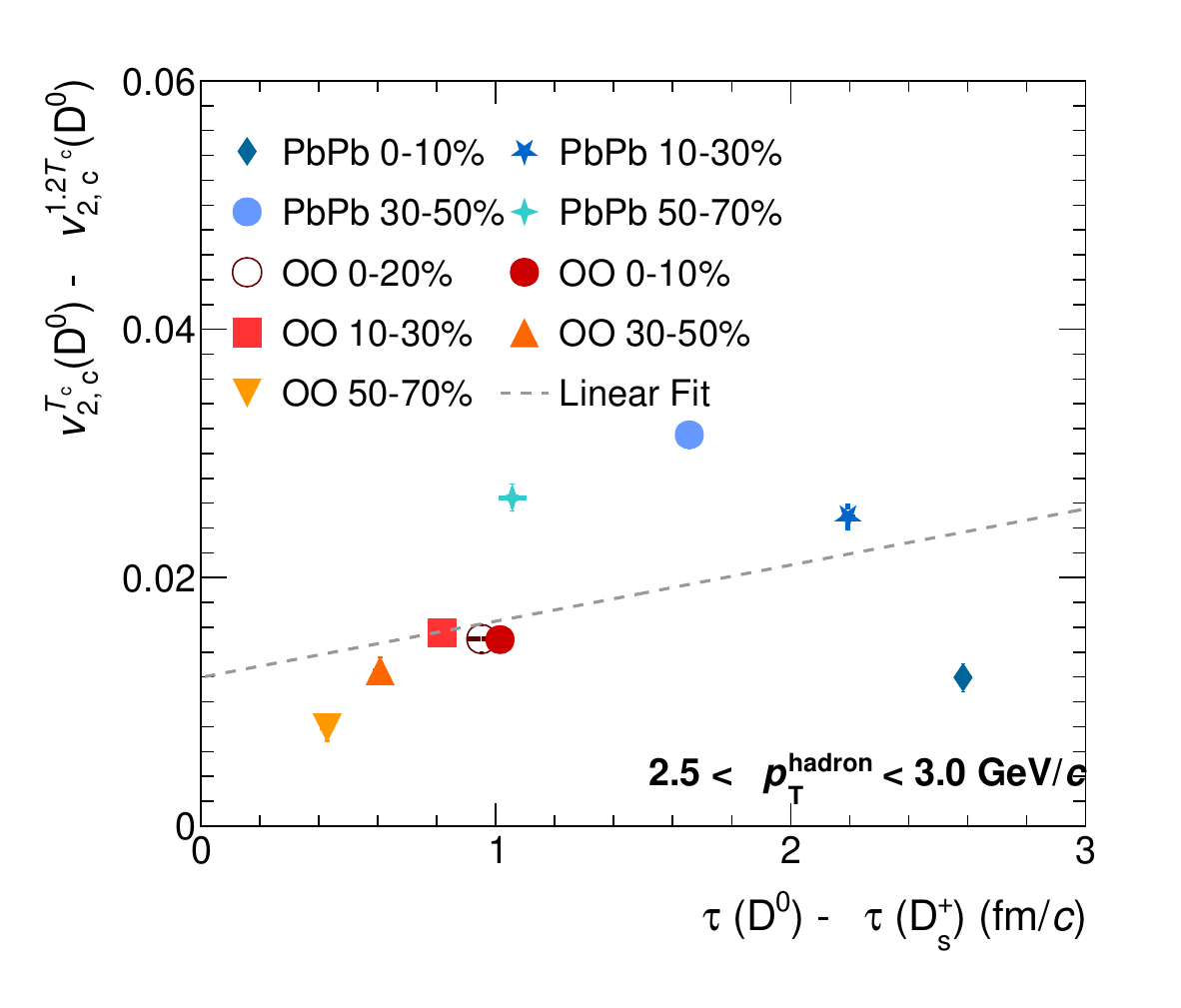}
    \caption{Partonic flow increment $\Delta v_{2,c}$ accumulated by parent charm quarks during $1.2\,T_c \to T_c$ as a function of $\Delta\tau_{\rm had}$, integrated over $2.5 < p_T^{\rm had} < 3.0$~GeV/$c$. Data points correspond to nine collision configurations---Pb--Pb 0--10\%, 10--30\%, 30--50\%, 50--70\% and O--O 0--10\%, 10--30\%, 30--50\%, 50--70\%, together with O--O 0--20\%---all evaluated under the sequential scenario.}
    \label{fig:7}
\end{figure}

\begin{figure}[t]
    \centering
    \includegraphics[width=1.08\linewidth]{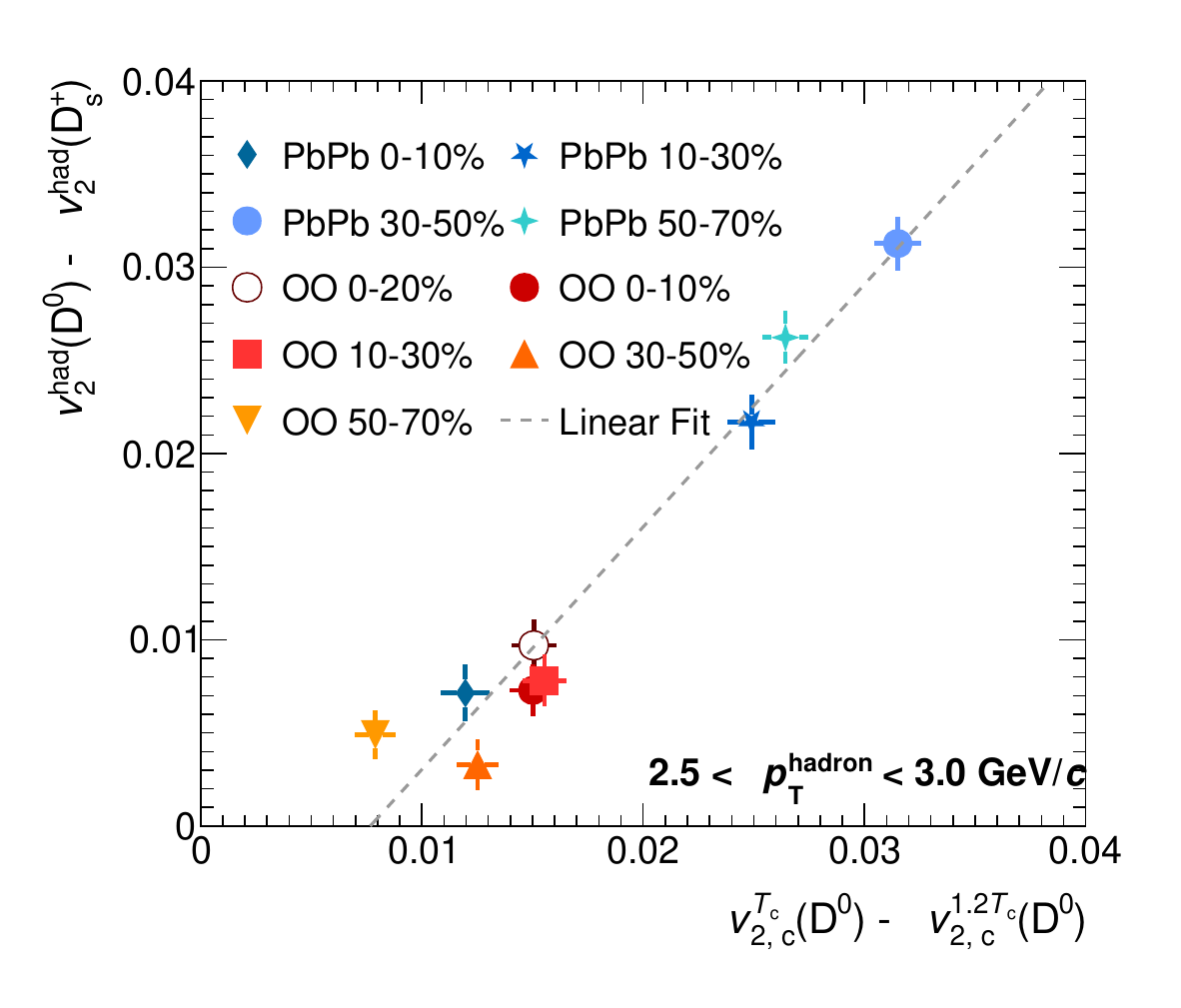}
    \caption{Hadronic flow splitting $\Delta v_2(D^0-D_s^+)$ versus partonic flow increment $\Delta v_{2,c}$, integrated over $2.5 < p_T^{\rm had} < 3.0$~GeV/$c$. Data points correspond to nine distinct collision configurations evaluated under the sequential scenario, spanning Pb--Pb 0--10\%, 10--30\%, 30--50\%, 50--70\% and O--O 0--10\%, 10--30\%, 30--50\%, 50--70\%, together with O--O 0--20\%.}
    \label{fig:8}
\end{figure}

Figure~\ref{fig:7} shows $\Delta v_{2,c}$ as a function of $\Delta\tau_{\rm had}$ for the nine collision configurations. As expected, $\Delta\tau_{\rm had}$ increases with system size and centrality: Pb--Pb collisions span $\Delta\tau_{\rm had} \sim 1.5$--$3$~fm/$c$, while all O--O configurations are concentrated below $1$~fm/$c$. Although a clear positive correlation is observed between $\Delta v_{2,c}$ and $\Delta\tau_{\rm had}$, the relationship is not strictly monotonic. The largest $\Delta v_{2,c}$ is found in Pb--Pb 30--50\% rather than in the most central 0--10\% bin, despite the latter having a longer $\Delta\tau_{\rm had}$. This reflects the critical interplay between the initial geometric eccentricity of the collision zone, which decreases toward central collisions, and the duration of the $1.2\,T_c \to T_c$ window, which increases with centrality. The 30--50\% class strikes an optimal balance between a sufficient initial spatial anisotropy and an adequate evolution time to convert it into momentum anisotropy. For the O--O systems, all configurations are confined to $\Delta\tau_{\rm had} < 1$~fm/$c$ and yield correspondingly small $\Delta v_{2,c}$. Here, the dynamics are dominated by the severely compressed time window: even though the initial eccentricity is larger in peripheral O--O collisions, the system lacks sufficient time to translate this spatial asymmetry into additional charm-quark flow. Consequently, the differences among the O--O configurations are marginal; $\Delta v_{2,c}$ in the 10--30\% bin is only slightly higher than in 0--10\% and 0--20\%, rather than following a monotonic trend with eccentricity. Ultimately, $\Delta v_{2,c}$ is governed not by the cooling time alone, but by the combined influence of the initial collision geometry and the late-stage QGP lifetime. As will be shown in Fig.~\ref{fig:8}, this partonic mapping is directly imprinted onto the final hadronic splitting.

Figure~\ref{fig:8} shows $\Delta v_2$ versus $\Delta v_{2,c}$ for all nine evaluated LHC configurations. The data points follow a clear linear trend. This linearity is a product of mixing the coalescence and fragmentation contributions across different collision configurations: the individual coal and frag channels for $D_s^+$ and $D^0$ are not strictly linear in $\Delta v_{2,c}$ when taken separately, but their superposition---sampled over these diverse systems with distinct coal-to-frag ratios---collapses onto a single line. Hadronic rescattering further amplifies the splitting on top of this baseline.

The collapse of all nine data points onto a single line, despite spanning widely different initial eccentricities and system sizes, establishes that $\Delta v_2$ is not controlled by the global collision history prior to $1.2\,T_c$, but is dictated entirely by the partonic anisotropy accumulated during the sequential hadronization window. We therefore propose the $D^0$--$D_s^+$ flow splitting as a ``hadronization chronometer'' of the QGP.

The O--O configurations occupy the small-$\Delta v_{2,c}$ limit of the universal line in Fig.~\ref{fig:8}, reflecting their severely compressed $1.2\,T_c \to T_c$ window; Pb--Pb systems extend the trend to correspondingly larger values. This unified behavior provides a natural explanation for the system-size ordering reported in Sec.~\ref{sec:results}: the substantially smaller $\Delta v_2$ observed in O--O 0--20\% relative to Pb--Pb 30--50\% is a direct consequence of O--O occupying the small-$\Delta v_{2,c}$ end of the universal line. On this line, O--O 0--20\% lies close to Pb--Pb 0--10\%, illustrating that a severely compressed $1.2\,T_c \to T_c$ window can place a small system in the same kinematic regime as the most central heavy-ion collisions. The optimal splitting is found in Pb--Pb 30--50\% rather than in the more peripheral 50--70\% bin, consistent with the competition between the initial geometric eccentricity $\varepsilon_2$ and the duration of the late-stage cooling window $\Delta\tau_{\rm had}$, as demonstrated in Fig.~\ref{fig:7}. Rather than signaling a breakdown of the sequential mechanism in small systems, the system-size ordering is a natural outcome of a universal mapping from the partonic to the hadronic sector. We emphasize that this scaling is established at $\sqrt{s_{NN}} = 5.36$~TeV and remains a phenomenological observation; whether it extends to different collision energies is an open question for future measurements.

\section{Summary}
In summary, we have investigated the elliptic-flow ($v_2$) splitting between $D^0$ and $D_s^+$ mesons using a Langevin transport model coupled to a hybrid coalescence-plus-fragmentation framework. We contrasted two hadronization paradigms---simultaneous versus sequential freeze-out---across collision configurations in O--O and Pb--Pb collisions.

For O--O 0--20\% collisions at $\sqrt{s_{NN}} = 5.36$~TeV, we presented theoretical predictions for both the elliptic flow and the particle composition. The sequential scenario reproduces the $v_2(D^0) > v_2(D_s^+)$ ordering observed in the preliminary ALICE data, whereas the simultaneous baseline yields the opposite ordering. The sequential scenario also predicts an enhanced $D_s^+/D^0$ yield ratio at low $p_T$ relative to the simultaneous case, and a systematic increase of this ratio from O--O to Pb--Pb, driven by the stronger partonic energy loss in the heavier system.

By decomposing the flow splitting into its partonic components, we showed that the simultaneous ordering arises from a phase-space selection effect in the coalescence channel, which cannot be reconciled with the experimental data even after hadronic rescattering. The sequential scenario reverses this ordering through the late-stage partonic flow accumulated by $D^0$-parent charm quarks during the $1.2\,T_c \to T_c$ interval.
In O--O collisions, the severe compression of the $1.2\,T_c \to T_c$ interval restricts the late-stage partonic $v_2$ buildup, potentially inducing an inverted ordering of $v_{2,c}(D_s^+) > v_{2,c}(D^0)$ at the parton level. Consequently, the hadronic rescattering phase---which imparts a larger flow increment to the $D^0$---proves critical for restoring the physical $v_2(D^0) > v_2(D_s^+)$ ordering in small collision systems.

To quantify this dynamical response, we performed a systematic scan across nine collision configurations spanning O--O and Pb--Pb, and identified a universal linear scaling between the final hadronic splitting $\Delta v_2$ and the partonic flow increment $\Delta v_{2,c}$. All nine configurations collapse onto the same line despite spanning widely different system sizes and initial eccentricities, demonstrating that $\Delta v_2$ is dictated by the partonic anisotropy accumulated during the $1.2\,T_c \to T_c$ window rather than by the global collision history. We further found that $\Delta v_{2,c}$ is governed by the competition between the initial geometric eccentricity and the duration of the $1.2\,T_c \to T_c$ window. In Pb--Pb, this competition selects the 30--50\% centrality class as the optimal balance; in O--O, the time window is so short that it dominates over eccentricity, confining all O--O configurations to the low-response end of the universal trend.

We therefore establish the $D^0$--$D_s^+$ flow splitting as a ``hadronization chronometer'' of the QGP at $\sqrt{s_{NN}} = 5.36$~TeV. O--O systems populate the low-response limit of this universal line, while Pb--Pb systems extend the trend to larger values, providing a natural explanation for the system-size ordering reported in this work. Whether this phenomenological scaling persists at different collision energies or with different ion species remains an open question for future measurements. 
\section{ACKNOWLEDGMENTS}
WD is supported by the National Key Research and Development Program of China (Grant No. 2024YFA1610804), BWZ is supported by a grant from the National Natural Science Foundation of China (Key Program) (No. 12535010), and JX is supported by the Helmholtz Research Academy Hesse for FAIR.
\providecommand{\noopsort}[1]{}\providecommand{\singleletter}[1]{#1}%

\end{document}